\documentclass[12pt]{article}
\usepackage{graphicx}
\usepackage{bm}
\usepackage{latexsym}
\newcommand{\beq}{\begin{equation}}
\newcommand{\eeq}{\end{equation}}
\newcommand{\bc}{\begin{center}}
\newcommand{\ec}{\end{center}}
\newcommand{\eeqa}{\end{eqnarray}}
\newcommand{\beqa}{\begin{eqnarray}}
\newcommand{\no}{\noindent}

\newcommand{\ra}{\rightarrow}
\newcommand{\na}{\nabla}

\newcommand{\al}{\alpha}
\newcommand{\be}{\beta}
\newcommand{\ga}{\gamma}
\newcommand{\Ga}{\Gamma}
\newcommand{\de}{\delta}

\newcommand{\et}{\eta}

\newcommand{\si}{\sigma}

\newcommand{\ph}{\phi}

\newcommand{\om}{\omega}
\newcommand{\ed}{\end{document} }
\begin{document}
\baselineskip=20pt
\title{Geometrical Origin of a Cosmological Term}

\author{Richard T. Hammond\thanks{rhammond@email.unc.edu }\\
Department of Physics\\
University of North Carolina at Chapel Hill\\
Chapel Hill, NC 27599
}
\date{\today}
\maketitle
\begin{abstract}
{Nonmetricity derived from a scalar field is shown to exist as a cosmic field, without direct coupling to matter. It leads to a variable cosmological term, a term that dominates the expansion in the early universe but dies away at later time.}

\end{abstract}

\section{Introduction}

The Dirac equation and General Relativity are two of the greatest achievements of the last century, and it did not take long to bring them together. In General Relativity, or in local Poincar\'e gauge generalizations \cite{hehl76}, the equations of gravitation may be written in terms of tetrads, and the Dirac equation may be written in the tangent space \cite{hammond02}. The precise coupling is obtained by assuming that the spinors transform according to the (usually the spin 1/2) representation of SO(1,3). Recently there has been growing interest in metric affine gravity, which includes torsion and a scalar field \cite{hehl95}. It is a local gauge theory based on the affine group SL(4R). An unsolved problem is the correct ``Dirac-type'' equation invariant under the affine group. The issue has been discussed in \cite{kirsch02}, and some believe it is not possible to properly write the Dirac equation in the case \cite{hehl05}.\footnote{The hubbub concerning the correct coupling in metric affine teleparallel theories\cite{obukhov} is not relevant here.}
     In this article I will sidestep this controversy and consider the Dirac equation in a Weyl integrable spacetime, and investigate some consequences of this coupling. This is not assumed to be a metric affine gauge theory, and  we assume that the Dirac equation,

\beq
\ga^\mu D_\mu\psi+im\psi=0
,\eeq
where $D_\mu=\partial_\mu+\Ga_\mu$, 
is invariant under the local Lorentz transformation

\beqa
\psi\ra S\psi=(1+\frac1 2\om_{ab}(x)\si^{ab})\psi \\ \nonumber
e_a^\mu\ra e_a^\mu+\om_a^{\ b}e_b^{\ \mu}
.\eeqa
 Details may be found in  Refs. \cite{hehl76} and \cite{hammond02}. The main point here is that we assume that the Dirac equation is invariant under the spin one half representation of $S(1,3)$,  so that
 
\beq
\Ga_\mu'=S\Ga_\mu S^{-1}-S_{,\mu}S^{-1}
\eeq
and
 
\beq\label{com}
[\si^{mn},\si^{ab}]=
\et^{na}\si^{mb}-\et ^{ma}\si^{nb}+\et ^{nb}\si^{am}-\et^{bm}
\si^{an}
,\eeq
where $\si^{ab}=(1/2)\ga^{[a}\ga^{b]}$.

It is well known that the correct form for $\Ga_\mu$ turns out to be
\beq\label{spincon}
\Ga_\mu=-\frac12\si_{ab}\Ga_\mu^{\ ab}
\eeq
where

\beq\label{sc1}
\Ga_{ab}^{\ \ c}=e^\al_{\ a}e^\be_{\ b}e_\si^{\ c}\Ga_{\al\be}^{\ \ \si}
-e^\al_{\ b}e_{\al\ ,a}^{\ c}
,\eeq
 and it is convenient to
define the quantity $\Ga_{\mu}^{\ ab}$ according to $\Ga_{\mu}^{\ ab}\equiv e_\mu^{\ n}\Ga_{n}^{\ ab}$. Latin indices are raised and lowered with $\eta_{ab}$, so for example $\Ga_{a}^{\ b c}=\et^{mb}\Ga_{am}^{\ \ \ c}$.

\section{Nonmetricity}

A cornerstone in the geometric framework that houses General Relativity is the
linear connection, expressed through the covariant derivative according to

\beq
D_\mu A^\si= A^\si_{\ ,\mu}+\Gamma_{\mu\nu}^{\ \ \si}A^\nu
.\eeq

\no The object $\Gamma_{\mu\nu}^{\ \ \si}$ is the linear connection, which was used in (\ref{sc1}), more commonly called the affine connection. The most general form can be written as

\beq\label{affine}
\Ga_{\mu\nu}^{\ \ \si}=\{_{\mu\nu}^{\ \si}\}
+S_{\mu\nu}^{\ \ \ \si}+S^\si_{\ \mu\nu}+S^\si_{\ \nu\mu}
+\frac1 2(Q_{\mu\nu}^{\ \ \ \si}+Q_{\nu\mu}^{\ \ \ \si}-Q^\si_{\ \mu\nu})
\eeq
where, on the right, we find the Christoffel symbol, 

\beq\label{christoffel}
\{_{\mu\nu}^{\ \si}\}=\frac1 2g^{\si\ga}
(g_{\ga\mu,\nu}+g_{\ga\nu,\mu}-g_{\mu\nu,\ga})
,\eeq
the torsion tensor,
\beq
S_{\mu\nu}^{\ \ \ \si}=\Ga_{[\mu\nu]}^{\ \ \ \si}
\eeq
and the nonmetricity tensor, $Q_{\si}^{\ \mu\nu}$, which is defined by

\beq\label{nm}
\na_\si g^{\mu\nu}\equiv Q_{\si}^{\ \mu\nu}
\eeq
where it is assumed that the metric tensor is symmetric.

Although it has been established in Refs.  \cite{hehl76} and \cite{hammond02} that the torsion is associated with the intrinsic spin of a particle, the interpretation of the nonmetricicity has not yet been able to enjoy such a definitive physical association. It was born with Weyl's struggle to geometrize electromagnetism, but the theory was found to be unphysical. The failure of Weyl's theory extended far beyond his approach, and seemed to taint the notion of nonmetricity in general. It has received a revival, though, in the metric affine local gauge theory, where it is associated with the dilaton field in \cite{hehl95}.
     
     From a geometric view, nonmetricity is associated with the change in length of a vector upon parallel transport. The failure of Weyl's theory stemmed from this change when it was associated with the electromagnetic potential. A more benign form assumes that the change in length vanishes upon parallel transport around a closed path, and is often called Weyl integrable geometry. In this case the nonmetricity cannot be associated with electromagnetism, and is written as

\beq\label{Q}
Q_{\si\mu\nu}=g_{\mu\nu}\ph_{,\si}
.\eeq

In the following we shall examine the coupling of this field to matter and investigate its role in cosmology, but this notion has attracted some attention in recent years. It was already shown in 1996 that cosmology in Weyl integrable spacetimes can be associated with a scalar field \cite{salim96}, \cite{oliveira97}, and nonsingular inflationary solutions were also found \cite{fabris98}. It was also proposed that the dilaton field associated with Weyl integrable geometry was related to the dark energy \cite{babourova04}. Weyl integrable geometry with torsion and quadratic Lagrangians has also been investigated \cite{puetzfeld01}. 
     
There are knotty issues that tangle the effectiveness of these theories. Often, there is an unknown function of the scalar field that is added in an ad hoc manner. Also, the scalar field can, in general, produce forces on particles, and these effects are often ignored. Associated with this is a careful study of implications of the Bianchi identity, which leads to the correct physical interpretation of mass, and the equation of motion. None of these issues arises here due to the noncoupling of nonmetricity to the Dirac particle.

The main result to be persued now is this: (\ref{spincon}) shows that $\Ga_\mu$, and therefore the covariant derivative, couples to the antisymmetric part of the spin connection, a well known result in metric theories. Thus,  (\ref{spincon}) with (\ref{affine}) shows that the nonmetricity does not couple to the spin connection. This has also been shown to be true from a spinor analysis.\cite{poberii} This means that the scalar field, as long as we stick to minimal coupling, does not couple to spin one half particles.\footnote{Others have adopted a ``Kosmann coupling,'' which includes a term proportional to the trace of the nonmetricity\cite{adak}.} What this boils down to is this: the scalar field will contribute to the curvature of space, but does not produce a direct force on particles.\footnote{
The scalar field does not couple to electromagnetism either. If we assume that the electromagnetic field, $F_{\mu\nu}$, is obtained from Minkowski space through minimal coupling, i.e., $\partial_\mu\ra D_\mu$, then, with (\ref{Q}), the nonmetricity does not couple to electromagnetism.}

It is important to note that same result (non-coupling of matter to the scalar field) occurs if, instead of using a Dirac source we examine a phenomenological source. In this case the Bianchi identities yield the equation of motion, and it will be seen that this also shows that the scalar field of nonmetricity exerts no forces on particles.\footnote{In fact, the nonmetricity acts like the cosmological term, and it will be seen that it can drive the universe through an expansive phase, and then die away.}

To see the details, consider the variational principle,

\beq\label{vp1}
\de(I_g+I_m)=0
\eeq
where\footnote{The curvature tensor is $
R_{\be\mu\nu}^{\ \ \ \ \si}=
\Ga_{\mu\nu ,\be}^{\ \ \si}-\Ga_{\be\nu ,\mu}^{\ \ \si}
+\Ga_{\be\ph}^{\ \ \ \si}\Ga_{\mu\nu}^{\ \ \ \ph}
-\Ga_{\mu\ph}^{\ \ \ \si}\Ga_{\be\nu}^{\ \ \ \ph}
$

and $R_{\mu\nu}=R_{\si\mu\nu}^{\ \ \ \ \si}$.}
\beq
I_g={1\over 16\pi}\int\sqrt{-g}Rd^4x
\eeq
and

\beq
\de I_m\equiv\frac12\int\sqrt{-g}T^{\mu\nu}\de g_{\mu\nu}d^4x
.\eeq
In the above, $R$ represents the full curvature scalar in non-Riemannian spacetime, with torsion set equal to zero, and the nonmetricity given by (\ref{Q}). Taking independent variations with respect to $g_{\mu\nu}$ and $\ph$ yields

\beq
G^{\mu\nu}+t^{\mu\nu}=8\pi T^{\mu\nu}
\eeq
where here $G^{\mu\nu}$ is the usual Einstein tensor in Riemannian space, $t^{\mu\nu}=(3/4)(g^{\mu\nu}\ph_{,\si}\ph^{,\si}-2\ph^{,\mu}\ph^{,\nu})$
and
\beq\label{box}
\Box\ph=0
.\eeq

The Bianchi identity gives $G^{\mu\nu}_{\ \ ;\nu}=0$ so that

\beq
8\pi T^{\mu\nu}_{\ \ ;\nu}=-\frac3 2\ph^{,\mu}\Box\ph
.\eeq
with \ref{box}, this shows that particles with no structure follow along geodesics and matter is conserved, which agrees with the Dirac coupling, which showed there is no interaction. Also, since there is no source, (\ref{box}) shows there are no 
$\ph\sim1/r$ type solutions.

\section{Cosmology}
     All these results suggest that $\ph$ should be interpreted as a cosmic field, and in an isotropic universe, that it should only be a function of time. In fact, assuming that 
     
\beq
ds^2=a^2(\eta)(d\eta^2-d\chi^2-\sin^2\chi d\theta^2-\sin^2\chi\sin^2\theta d\ph^2)
\eeq
(\ref{box}) gives

\beq
\ph '\equiv \ph_{,\eta}={2A\over a^2}
\eeq
where $A$ is a constant and

\beq\label{G0eta}
G_0^{\ 0}=8\pi T_0^{\ 0} +{3A^2\over a^4}
.\eeq

Alternatively, one may use $ds^2=dt^2 -(1-kr^2)^{-1}dr^2-r^2d\theta^2-r^2\sin^2\theta d\ph^2$
where $\dot\ph=\ph_{,t}={2A^2}/a^3$ and (\ref{G0eta}) becomes

\beq\label{adot}
\dot a^2+k=\frac L a +{A^4\over a^4}
\eeq
where $L=4M/(3\pi)$, the constant $M$ is given by $\rho=M/(2\pi^2 a^3)$, 
$T_0^{\ 0}=\rho$, and $k=1,-1,0$ for a closed, open, and flat universe.

Equation (\ref{adot}) shows that for large $a$ the effect of the scalar field dies away, but for small $a$ dominates over the matter terms. This shows that the scalar field of nonmetricity may be viewed as a variable cosmological term. It explains in a natural, i.e., geometrical, fashion why the cosmological term is dominant in the early universe, and why it decays to zero in the late universe. 
Only when
$a\sim\sqrt[3]{A^4/L}$ does matter weigh in.
     
As a simple case let us consider a spatially flat universe. Let us compare the solution when only matter is present and call this $a_\rho$ to the solution when only the scalar field is present (early universe),  $a_\rho$. For these special cases (\ref{adot}) may be integrated, from which we see,

\beq
\left(a_\phi\over a_\rho\right)^3\sim 1/t,
\eeq
showing that at early times the scalar field drives the expansion of the universe, and that it fades away at later times. In this view, the nonmetricity acts as a helping hand, expanding the universe when it was a baby and, like a good parent, steps away as the universe ages. Future work will examine positive and negatively curved spatial geometry, as well as the general problem of obtaining solution with both matter and the scalar field present.

\ed
\begin{thebibliography}{}



\bibitem{hehl76} F. W. Hehl, P. von der Heyde, G. D. Kerlick, and J. M. Nester 
 {\it Rev. Mod. Phys.} {\bf 48} 393 (1976).

\bibitem{hammond02} R. T. Hammond,  Rep. Prog. Phys. {\bf 65}, 599 (2002), see also Ref. \cite{hehl76}.

\bibitem{hehl95} F.W. Hehl, J. D. McCrea, E. W. Mielke, Y. Ne'eman, {\it Phys. Rep.}
{\bf 258}, 1-171 (1995).


\bibitem{kirsch02} I. Kirsch and D. Sijacki, Class. Quantum Grav. {\bf 19}, 3157 (2002).

\bibitem{hehl05} F. W. Hehl, private communication.

\bibitem{obukhov}Y. N. Obukhov and J. G. Pereira, Phys. Rev. D {\bf 67}, 044016 (2003);
{\bf 69} 128502 (2004); E. W. Mielke, Phys. Rev. D {\bf 69}, 128501; J. W. Maluf, Phys. Rev. D {\bf 67}, 108501 (2003).


\bibitem{poberii} E. A. Poberii, ``Spinor-type objects in metric-affine spacetime,'' JHEP07(1998)016.

\bibitem{adak} M. Adak, T. Dereli, and L. Ryder, Phys. Rev. D {\bf 69}, 123002 (2004); see also gr-qc/0208042 (2002).













\bibitem{salim96} J. M. Salim, and S. L. Sautu, Class. Quantum Grav. {\bf 13}, 353 (1996).

\bibitem{oliveira97} H . P. de Oliveira, J. M. Salim, and S. L. Sautu, Class. Quantum Grav. {\bf 14}, 2833 (1997).

\bibitem{fabris98} J. C. Fabris, J. M. Salim, and S. L. Sautu, Mod. Phys. Lett. A {\bf 13} 953 (1998).

\bibitem{babourova04} O. Barbourova, Gravitation \& Cosmology {\bf 10} 121, (2004),
gr-qc/0507104.
   
\bibitem{puetzfeld01} D. Puetzfeld and R. Tresguerres,  
Class. Quantum Grav. {\bf 18}, 677 (2001); D. Puetzfeld,  
Class. Quantum Grav. {\bf 19}, 3263 (2002);gr-qc/0402026. 
   
   
   
 \bibitem{hammond99}  R. T. Hammond,
Gen. Rel.  Grav. {\bf 31} 889, 1999.
     
   
\bibitem{vishwakarma05} R. G. Viswakarma, Gen. Rel.  Grav. {\bf 37} 1305, 2005.
  

\bibitem{amendola05} L. Amendola and F. Finelli, Phys. Rev. Lett. {\bf 94}, 221303 (2005).







\end{thebibliography}
